# Domain-wall motion induced by spin transfer torque delivered by helicity-dependent femtosecond laser

Boyu Zhang,[1,2,3,*] Yong Xu,[1,2,*] Weisheng Zhao,[1,4,†] Daoqian Zhu,[1] Huaiwen Yang,[1,4] Xiaoyang Lin,[1] Michel Hehn,[2] Gregory Malinowski,[2] Nicolas Vernier,[3] Dafiné Ravelosona,[3] Stéphane Mangin[2]

[1]Fert Beijing Institute, BDBC, School of Microelectronics, Beihang University, Beijing 100191, China.
[2]Institut Jean Lamour, CNRS UMR 7198, Université de Lorraine, 54000 Nancy, France.
[3]Centre de Nanoscience et de Nanotechnologie, CNRS UMR 9001, Université Paris-Sud, Université Paris-Saclay, 91400 Orsay, France.
[4]Beihang-Goertek Joint Microelectronics Institute, Qingdao Research Institute, Beihang University, Qingdao 266000, China
[*]These authors contributed equally to this work.
[†]Corresponding author. Email: weisheng.zhao@buaa.edu.cn (W.Z.)

In magnetic wires with perpendicular anisotropy, moving domain with only current or only circularly polarized light requires a high power. Here, we propose to reduce it by using both short current pulses and femtosecond laser pulses simultaneously. The wires were made out of perpendicularly magnetized film of Pt/Co/Ni/Co/Pt. The displacement of the domain wall is found to be dependent on the laser helicity. Based on a quantitative analysis of the current-induced domain wall motion, the spin orbit torque contribution can be neglected when compared to the spin transfer torque contribution. The effective field of the spin transfer torque is extracted from the pulsed field domain wall measurements. Finally, our result can be described using the Fatuzzo-Labrune model and considering the effective field due to the polarized laser beam, the effective field due to spin transfer torque, and the Gaussian temperature distribution of the laser spot.

# I. INTRODUCTION

Efficient control of the domain wall (DW) motion is essential for the development of high-performance racetrack memories [1-3]. Spin transfer torque (STT) [4,5] has been widely used to manipulate DWs. However, STT-driven DW motion often requires a high current density up to $10^7$ A/cm$^2$ [6] or an external perpendicular magnetic field [7], which makes this technique difficult to implement for applications.

Recently, all-optical helicity-dependent switching (AO-HDS) has been reported for a variety of ferromagnetic materials with perpendicular magnetic anisotropy (PMA) [8-13]. Circularly polarized femtosecond (fs) laser pulses are shown to deterministically switch the magnetization up or down depending on the laser helicity. Moreover, fs laser provides a new way for controlling DW motion, where all optical helicity-dependent DW motion driven by fs laser pulses was demonstrated [14,15].

[Co/Ni] multilayers have been reported to have a high spin polarization and their PMA can be easily tuned [16-18]. Furthermore, [Co/Ni] multilayers are widely used for current-induced DW motion [6,7,19-20] and AO-HDS has also been reported in [Co/Ni] [9,10], which makes [Co/Ni] an ideal system for studying the combined effect of STT and AO-HDS. In this paper, we study the STT-driven DW motion assisted by fs laser pulses in perpendicularly magnetized Pt/Co/Ni/Co/Pt thin films.

The paper is organized as follows: in Sec. II, we characterize the magnetic properties of the sample, and show the combined effect of current pulses and laser pulses; the contributions of spin orbit torque (SOT) and STT are quantified in Sec. III-A and III-B, respectively; in Sec. III-C, we present an analysis based on the Fatuzzo-Labrune model [21,22] to reproduce the experimental results in Sec. II.

## II. RESULTS

The investigated sample is a sputtered thin film namely Ta(3 nm)/Pt(3 nm)/Co(0.3 nm)/Ni(0.6 nm)/Co(0.3 nm)/Pt(3 nm) deposited on a glass substrate. Symmetric Pt (3 nm) layers were grown beneath and above Co/Ni/Co in order to minimize the Dzyaloshinskii–Moriya interaction (DMI). The magnetic properties were characterized by a superconducting quantum interference device-vibration sample magnetometer (SQUID-VSM) at room temperature. The sample shows strong PMA [Fig. 1(a)], where a coercivity field $H_c$ of 25 Oe and a saturation magnetization $M_S$ of 770 emu/cm$^3$ are measured.

The effective DMI field $H_{DMI}$ was obtained through the method of asymmetric expansion of domain bubbles in the creep regime using a Kerr microscope described in reference [23]. Fig. 1(b) shows the velocities of the DW propagation with an out-of-plane field and an in-plane field ($H_{inplane}$) applied simultaneously. For a fixed out-of-plane field of 60 Oe, the minimum DW velocity appears at $|H_{inplane}|$=100 Oe, which suggests an effective DMI field $H_{DMI}$ of +100 Oe.

The sample was then patterned into wires of 4 to 10 μm width by UV optical lithography. To study the AO-HDS effect, a laser beam (35 fs pulse duration, 800 nm wavelength, 5 kHz repetition rate) was swept over the wire. Fig. 1(c) shows that AO-HDS is observed for the 4 μm Co/Ni/Co wire with a laser fluence of 9 mJ/cm$^2$. The study of the fluence dependence shows that the minimal fluence for observing AO-HDS in such condition is 4.5 mJ/cm$^2$, which corresponds to the threshold fluence. In the following, laser fluences below the threshold were used to assist the current-driven DW motion.

The current pulses were synchronized with the 5-kHz-repetition-rate laser pulses. The inset of Fig. 1(d) displays the laser intensity as a function of the distance from the laser beam center. The beam diameter (Full width at half maximum, FWHM) is 47 μm. The center of the laser spot, indicated by the star in Fig. 1(d), was placed at 8 μm away from the DW. A single DW was nucleated in the wire before the experiments.

To illustrate the influence of the fs laser, the DW was driven by μs current pulses with and without the assistance of laser pulses [Fig. 1(d)]. Current pulses with a 10 μs pulse duration and a 5 kHz frequency were applied across the 4 μm wire. Without the assistance of the fs laser pulses, it is confirmed that μs current pulses with a current density lower than $16.5 \times 10^6$ A/cm$^2$ cannot move the DW (See III-B for more discussions). Fig. 1(d) shows DW motion assisted by left-circularly (σ-) polarized, linear (L) polarized, and right-circularly (σ+) polarized laser pulses. The Kerr images were taken after applying both current pulses of $7.3 \times 10^6$ A/cm$^2$ and laser pulses of 4 mJ/cm$^2$ for 10 seconds. The right-circularly (σ+) laser beam favors a large DW displacement as the DW moves further than the center of the beam, while linear (L) polarized laser beam induces a moderate DW displacement towards the beam center. In contrast, the left-circularly (σ-) polarized laser beam pins the DW.

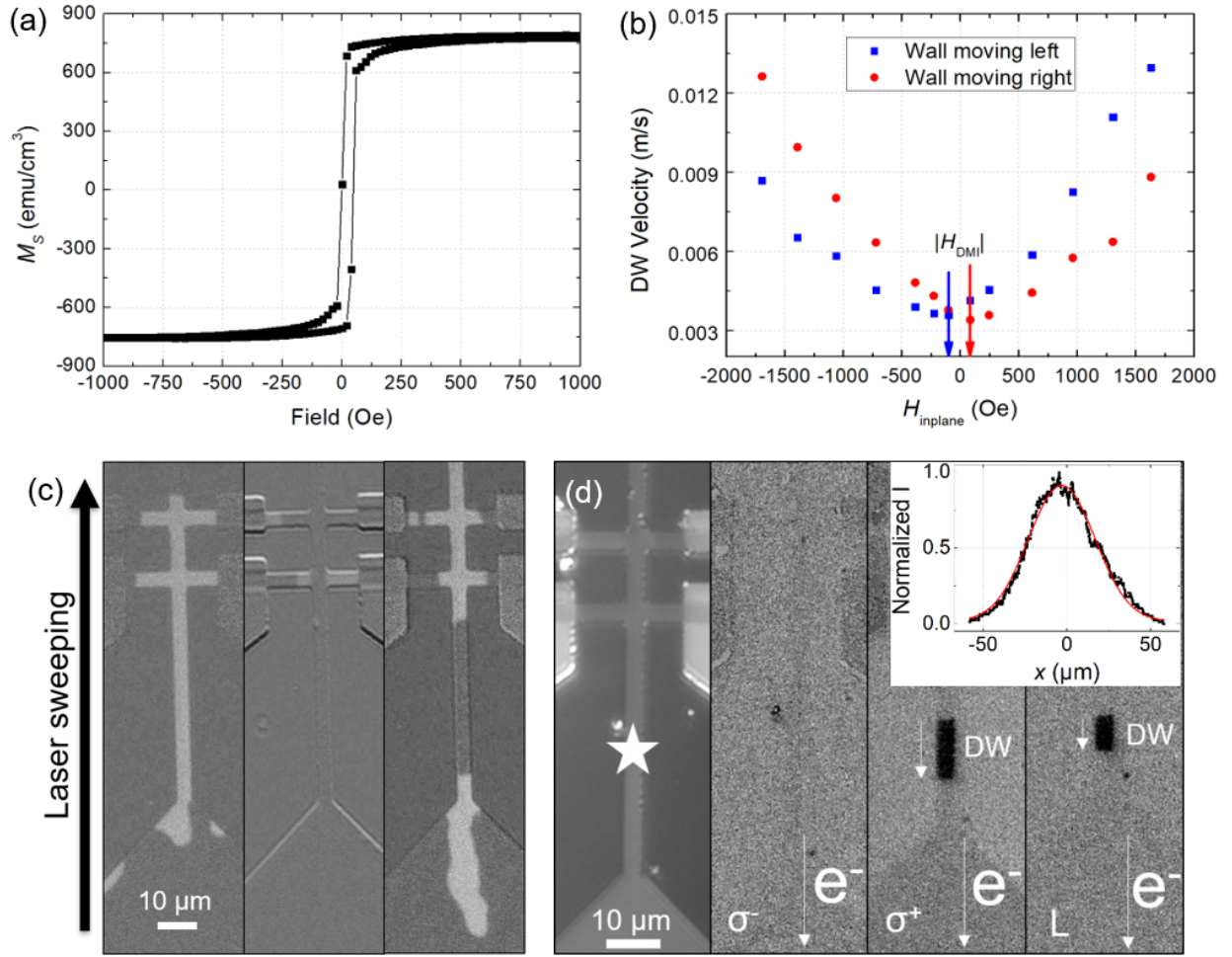

FIG. 1. Measurements on Ta(3 nm)/Pt(3 nm)/Co(0.3 nm)/Ni(0.6 nm)/Co(0.3 nm)/Pt(3 nm). (a) Out-of-plane hysteresis loop obtained by measuring the magnetization as a function of the field applied perpendicular to the thin film. (b) Field-induced DW propagation velocities in the creep regime as a function of the in-plane magnetic field $H_{inplane}$. Positive $H_{inplane}$ is defined as pointing to the right. The out-of-plane field is fixed at 60 Oe with the direction out of the thin film. (c) Kerr images of the 4 μm wire (patterned from the thin film) after linear (L), right-circularly (σ+) and left-circularly (σ-) polarized laser beams were swept over the wire with a fluence of 9 mJ/cm$^2$. The initial magnetization saturation direction is M↑, and the white contrast corresponds to a reversal to M↓. (d) Kerr images of the 4 μm wire after linear (L), right-circularly (σ+) and left-circularly (σ-) polarized laser beams shined at a fixed point indicated by the star (8 μm away from the DW) on the wire with a fluence of 4 mJ/cm$^2$ below the AO-HDS threshold for 10 seconds, together with synchronized current pulses of 7.3×10$^6$ A/cm$^2$ and 10 μs pulse duration (5 kHz repetition rate). The initial magnetization saturation direction is M↓, and the black contrast corresponds to a reversal to M↑. Inset: Gaussian profile of the laser spot with a FWHM of 47 μm.

## III. DISCUSSION

### A. The SOT contribution

The second harmonic technique described in Ref. [24] was used to quantify the SOT contribution in our sample. The inset of Fig. 2(a) shows the optical image of the 10 μm Hall bar as well as the experimental geometry. A sinusoidal current with a peak value of $I_{ac}$ oscillating at $f$=133.33 Hz was injected along the $x$-direction, while the first harmonic voltage $V_\omega$ and second harmonic voltage $V_{2\omega}$ were measured along the $y$-direction with lock-in amplifiers. The in-plane magnetic field $H_{x(y)}$ was applied along the $x$- or $y$-direction. The spin Hall angle $\theta_{SH}$, the damping-like field $\Delta H_x$, and the field-like effective field $\Delta H_y$ can be deduced from the following equation:

$$\Delta H_{x(y)} = -2 \frac{\partial V_{2\omega}}{\partial H_{x(y)}} / \frac{\partial^2 V_\omega}{\partial H_{x(y)}^2} \qquad (1)$$

$$\theta_{SH} = -\frac{2eM_s t_f}{\hbar} \times \frac{\Delta H_x}{J_{ac}} \qquad (2)$$

where $e$ is the electron charge, and $t_f$ is the thickness of the ferromagnetic layer. By parabolic fitting of the first harmonic signal and linear fitting of the second harmonic signal, $\partial V_{2\omega}/\partial H_{x(y)}$ and $\partial^2 V_\omega/\partial H^2_{x(y)}$ can be obtained, and $\Delta H_{x(y)}$ can be deduced from Eq. (1). In Fig. 2(a), the $\theta_{SH}$ was calculated from the damping-like effective fields $\Delta H_x$ with Eq. (2) and plotted as a function of $J_{ac}$, considering $J_{ac}=I_{ac}/(t\times w)$ with sample thickness $t$=10.2 nm, width $w$=10 μm and the sinusoidal current with a peak value of $I_{ac}$. Although the thicknesses are the same for the top and bottom Pt layers, we still have an average $\theta_{SH}$ of 0.04 which may be due to the different bottom Pt/Co and top Co/Pt interfaces. Fig. 2(b) shows the effective field $\Delta H_{x(y)}$ as a function of different $J_{ac}$, where the effective damping-like torque efficiency $\Delta H_x/J_{ac}$=1.3 Oe/($10^6$ A/cm$^2$) and field-like torque efficiency $\Delta H_y/J_{ac}$=0.13 Oe/($10^6$ A/cm$^2$) were extracted by linear fitting of the curve. The field-like torque induces a negligible Rashba effect, which is consistent with the results for Pt/Co/Pt structure [25].

As the effective DMI field of +100 Oe is not sufficient to favor a left-handed chiral Néel DWs [26], the SOT with a $\theta_{SH}$ of 0.04 and $\Delta H_x/J_{ac}$ of 1.3 Oe/($10^6$ A/cm$^2$) cannot induce DW motion with only current pulses, but it may favor the DW motion against the electron flow [27]. However, our experiment in Fig. 1(d) shows that the DW propagates in the direction of the electron flow, indicating that the SOT contribution for the DW motion is negligible compare to the STT contribution.

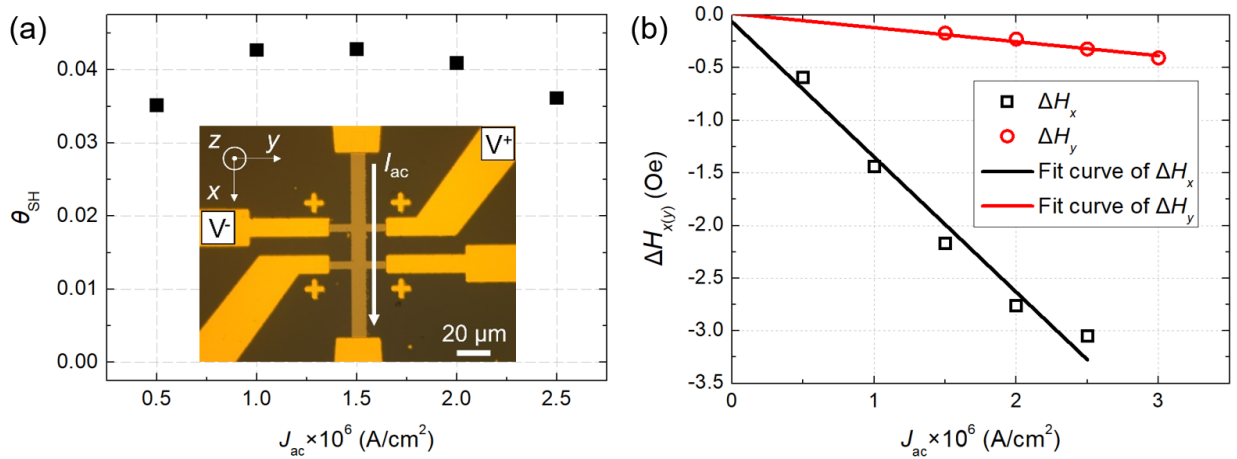

FIG. 2. Measurements on Ta(3 nm)/Pt(3 nm)/Co(0.3 nm)/Ni(0.6 nm)/Co(0.3 nm)/Pt(3 nm). (a) The spin Hall angle $\theta_{SH}$ as a function of $J_{ac}$ using the second harmonic technique. The inset

shows the microscope image of the 10 μm Hall bar and the schematic illustration of the second harmonic experimental geometry. (b) The damping-like effective field $\Delta H_x$ and field-like effective field $\Delta H_y$ deduced from the above measurements as a function of $J_{ac}$. The open symbols were extracted by fitting the first and second harmonic signals, and the solid curves correspond to linear fittings. The torque efficiency is given by the slope of the linear fitting $\Delta H_{x(y)}/J_{ac}$.

## B. The STT Contribution

To quantify the STT contribution, we performed the experiment with the μs field pulses applied perpendicular to the sample. Field and current pulses were synchronized with both 2 μs pulse duration and 0.5 μs delay. After the nucleation of a single DW in the wire, $7.3 \times 10^6$ A/cm$^2$ or $14.6 \times 10^6$ A/cm$^2$ current pulses were injected synchronously with different field pulses. DW displacement after each synchronized pulse was measured by the image difference technique. The DW velocity is calculated by dividing the DW displacement by 2 μs (the pulse duration). Fig. 3(a) displays the DW velocity plotted as a function of the magnetic field $H_{pulse}$. The black curve with square symbol shows the field-driven DW motion without current pulses. Both positive and negative currents increase the DW velocity, while the increase is more significant for negative currents.

In order to estimate the contribution from STT, $\Delta H$ is introduced as the average field shift compared to the curve without current pulses [7]. Figure 3(b) shows $\Delta H$ as a function of current density $J$ and the fitted curve using $\Delta H = +\varepsilon J + \eta J^2$, where $\varepsilon = -0.6$ Oe/($10^6$ A/cm$^2$) and $\eta = 0.29$ Oe/($10^6$ A/cm$^2$)$^2$. Both adiabatic and nonadiabatic components may play a significant role in the DW motion in the investigated wire. The nonadiabatic term proportional to the current density ($\varepsilon J$) acts as an effective field of STT $H_{STT}$ [28]. The quadratic adiabatic term ($\eta J^2$) may be related to Joule heating, where it contributes to an increase of temperature by $T + \delta J^2$ with a constant $\delta$. The temperature rise due to Joule heating of current pulses can be estimated by the following equation [29]:

$$\Delta T = \frac{RI^2 \times \{\ln[16K/dCw^2] + \ln(\tau_{pulse})\}}{2\pi l K} \quad (3)$$

where the specific heat $C=750$ J kg$^{-1}$ K$^{-1}$, thermal conduction $K=1.4$ W m$^{-1}$ K$^{-1}$, density $d=2500$ kg/m$^3$ for the glass substrate, $R=1350$ Ω, $\omega=4$ μm, $l=94$ μm for the wire, and the pulse duration $\tau_{pulse}=2$ μs. With the current density between $7.3\times10^6$ A/cm$^2$ and $14.6\times10^6$ A/cm$^2$, the temperature rise from an injected current pulse is around 5.75 K-23 K.

Therefore, the temperature rise due to Joule heating may contribute to the increase of DW velocity as Joule heating helps the DW to overcome the pinning energy barrier [30], while the velocity increases more in the negative current case as the negative current gives a positive $H_{STT}$.

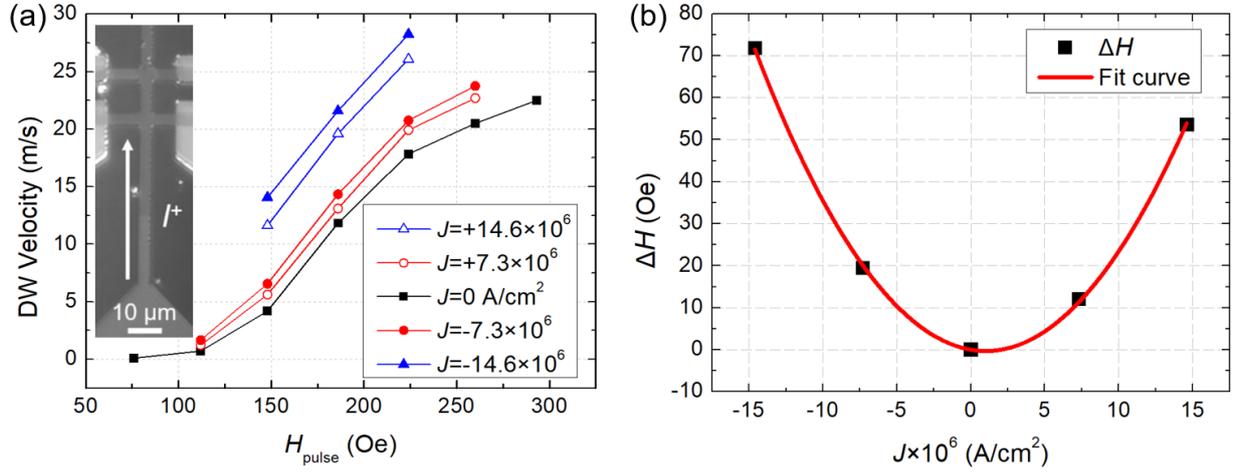

FIG. 3. Measurements on Ta(3 nm)/Pt(3 nm)/Co(0.3 nm)/Ni(0.6 nm)/Co(0.3 nm)/Pt(3 nm). (a) DW velocity as a function of the magnetic field $H_{pulse}$ for different current pulses. Current pulses of $7.3\times10^6$ A/cm$^2$ or $14.6\times10^6$ A/cm$^2$ were injected synchronously with the magnetic field pulses, where the pulse duration is both 2 μs with 0.5 μs delay. (b) $\Delta H$ as a function of current density $J$, where $\Delta H$ is the average field shift compared to the curve without current pulses. $\Delta H$ is positive if the shift is to the left. The black symbols were extracted from the experimental data and the red curve was fitted by $\Delta H = +\varepsilon J + \eta J^2$.

## C. Modeling

With the SOT contribution excluded and the STT contribution estimated, it is now possible to model the combined effect of STT and the fs laser beam. For the circularly polarized laser beam, the roles of the temperature and the effective field need to be distinguished. As the DW displacement induced by right-circularly (σ+) polarized laser beam exceeds the beam center and left-circularly (σ-) polarized laser beam doesn't induce any DW motion in Fig 1, the laser helicity can be interpreted as an effective field $H_\sigma$ and the direction of $H_\sigma$ depends on the laser helicity. $H_\sigma$ can be explained by the inverse Faraday effect (IFE) [14,31], where σ− favors a reversal to M↓ and σ+ favors a reversal to M↑. Magnetic circular dichroism (MCD) can also induce an effective field. As σ- is more absorbed by the M↑ domain and σ+ is more absorbed by the M↓ domain [14,32], a difference in absorption of M↑ and M↓ domains results in an additional temperature gradient and an effective field of MCD across the DW, where σ- favors a propagation towards M↑ domain and σ+ favors a propagation towards M↓ domain [14,33]. For linear (L) polarized laser beam without IFE or MCD, our results in Fig. 1 show that the DW propagates towards the hot region at the beam center due to the laser temperature distribution [inset of Fig. 1(d)].

We use the Fatuzzo-Labrune model [21,22] to evaluate the DW velocity under the combined action of the laser pulses and the current pulses:

$$v = v_0 \exp(-\frac{E - 2H_{\text{eff}} M_S V_B}{k_B T}) \qquad (4)$$

where $E$ represents the pinning energy barrier to be overcome in order to enable the DW motion within the Barkhausen volume $V_B$, $H_{\text{eff}} = H_\sigma + H_{\text{STT}}$ is an effective field that contains the contribution from helicity-dependent fs laser and STT, and $H_{\text{STT}} = \varepsilon J$.

The laser beam can be regarded as a Gaussian distribution of effective field $H_\sigma$ and temperature $T$, where $H_\sigma$ and $T$ represent the contribution from laser helicity and laser temperature distribution, respectively. We assume that the center of the laser beam gives a maximum temperature of 600 K [14] and a $|H_\sigma|$ of 3 Oe. A current density $J$ of $7.3 \times 10^6$ A/cm$^2$ gives a $|H_{\text{STT}}|$ of 4.38 Oe and a temperature rise of 29 K to $T$ according to Eq. (3) with the pulse duration $\tau_{\text{pulse}}$ of 10 μs. As the DW velocity $v$ is a function of the Gaussian distribution of $H_\sigma$ and $T$ related to the laser position $x$ with $v = dx/dt = f(x)$, by solving the equation, the DW displacement $x$ as a function of the time $t$ can then be obtained with $V_B = 10^{-23}$ m$^{-3}$, $v_0 = 2 \times 10^{12}$ μm/s, as shown in Fig. 4.

The σ+, linear, and σ- beams induce large, moderate, and small DW displacements, respectively. The simulation results are in agreement with the experiment in Fig. 1(d). Based on those parameters, implementing only current pulses or laser pulses into the model gives a vanishing DW velocity, which explains why only the current pulses or laser pulses (below AOS threshold) cannot drive the DW motion in our system.

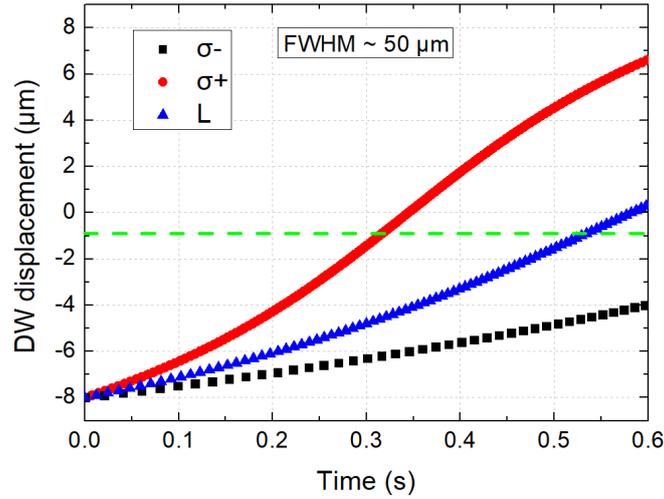

FIG. 4. Time-dependent simulations of the DW displacement based on the Fatuzzo-Labrune model. The DW motion is induced by synchronized current pulses and linear (L), right-circularly (σ+) or left-circularly (σ-) polarized laser pulses with a $H_{STT}$ of 4.38 Oe, a Gaussian distribution of $H_σ$ and $T$, with a FWHM of ~50 μm. The green dashed line corresponds to the position of the laser beam center.

## IV. CONCLUSION

To conclude, we have experimentally investigated the combined effect of STT and fs laser in perpendicularly magnetized 4 μm Pt/Co/Ni/Co/Pt wires. We demonstrate that circularly polarized laser pulses propagate or pin the DW depending on the laser helicity. Current pulses as low as $7.3×10^6$ A/cm² is sufficient for DW motion without any magnetic field. Through the asymmetric field-driven domain expansion and the second harmonic measurements, we exclude the SOT contribution as the sample has a small effective DMI field $H_{DMI}$ of +100 Oe and a spin Hall angle of +0.04. The effective field due to STT, which is the dominate contribution, is analyzed by pulsed-field DW measurements. The combined effect of STT and fs laser beam is explained with a model considering STT and helicity-dependent optical effect as effective fields, and the temperature distribution. Our approach highlights a new path for efficient DW manipulation, which can enable the development of new families of nanodevices combining photonics and spintronics.

## ACKNOWLEDGMENTS

The authors thank the fruitful discussions with Prof. Eric Fullerton. The authors gratefully acknowledge the National Natural Science Foundation of China (Grant No. 61627813, 61571023, 51602013), Program of Introducing Talents of Discipline to Universities (No. B16001), the National Key Technology Program of China (No. 2017ZX01032101), Young Elite Scientists Sponsorship Program by CAST (No. 2018QNRC001) and the China Scholarship Council. This work was supported by the ANR-15-CE24-0009 UMAMI and by the ANR-Labcom Project LSTNM, by the Institut Carnot ICEEL for the project « Optic-switch » and Matelas, and by the French PIA project 'Lorraine Université d'Excellence', reference ANR-15-IDEX-04-LUE. Experiments were performed using equipment from the TUBE. Davm funded by FEDER (EU), ANR, Région Grand Est and Metropole Grand Nancy.